\documentclass[letterpaper, 10 pt, conference]{ieeeconf}  

\IEEEoverridecommandlockouts                              

\usepackage{graphics} 
\usepackage{epsfig} 
\usepackage{mathptmx} 
\usepackage{times} 
\usepackage{amssymb}  

\usepackage{latexsym}
\usepackage{amsmath,amsfonts,paralist}

\usepackage{algpseudocode,algorithm,algorithmicx}
\usepackage{graphicx}
\usepackage{float}

\usepackage{bbold}
\usepackage{flushend}

   \let\g=\gamma  \let\d=\delta \let\e=\varepsilon
     \let\l=\lambda

  \let\r=\rho 

\let\bs=\backslash

\def\ul{\underline}

\def\ol{\overline}

\def\to{\rightarrow}

\def\<{\langle}
\def\>{\rangle}
\def\(({\left(}
\def\)){\right)}
\def\[[{\left[}
\def\]]{\right]}

\newcommand{\beq}{\begin{equation}}
\newcommand{\eeq}{\end{equation}}

\newcommand{\Tr}{\text{Tr}}

\title{\LARGE \bf
Spectral Detection on Sparse Hypergraphs
}

\author{Maria Chiara Angelini$^{1}$, Francesco Caltagirone$^{2}$, Florent Krzakala$^{2,3}$ and Lenka Zdeborov\'a$^{4}$
\thanks{*The research leading to these results has received funding from the
European Research Council under the European Union's $7^{th}$
Framework Programme (FP/2007-2013)/ERC Grant Agreement 307087-SPARCS.}
  \thanks{$^{1}$Dipartimento di Fisica, Sapienza Universit\'a di Roma,
    P.le Aldo Moro 2, 00185 Roma, Italy}%
  \thanks{$^{2}$PSL Research University, Laboratoire de Physique
    Statistique, UMR CNRS \& {\'E}cole Normale Sup{\'e}rieure, 24 Rue
    Lhomond, Paris, France}%
  \thanks{$^{3}$Sorbonne Universit\'es, UPMC Univ Paris 06,
    Laboratoire de Physique Statistique, 24 Rue Lhomond, Paris,
    France}%
  \thanks{$^{4}$Institut de Physique Th\'eorique, CEA Saclay, 91191
    Gif-sur-Yvette, France}%
}

\begin{document}

\maketitle
\thispagestyle{empty}
\pagestyle{empty}

\begin{abstract}
  We consider the problem of the assignment of nodes into communities
  from a set of hyperedges, where every hyperedge is a noisy
  observation of the community assignment of the adjacent nodes.  We
  focus in particular on the sparse regime where the number of edges
  is of the same order as the number of vertices.
  We propose a spectral method based on a generalization of the
  non-backtracking Hashimoto matrix into hypergraphs.  We analyze its
  performance on a planted generative model and compare it with other
  spectral methods and with Bayesian belief propagation (which
  was conjectured to be asymptotically optimal for this model).  We
  conclude that the proposed spectral method detects communities
  whenever belief propagation does, while having the important
  advantages to be simpler, entirely nonparametric, and to be able to
  learn the rule according to which the hyperedges were generated
  without prior information.
\end{abstract}

\section{Introduction}
Detecting information about the vertex properties that is hidden (or
encoded) in the structure of a graph is a central issue in many
problems in physics, biology and computer science. In fact, many
systems of interest are composed of a large number of variables about
which we do not have any information but the relationships (or part of
the relationships) between them. Starting from this knowledge we
aim at inferring individual properties of the nodes. 

In this context, the main approaches are statistical inference, where
the detection is based on the assumption of a generative model for the
graph \cite{Holland83,wangwong}, and spectral methods \cite{luxburg}.
For some classes of generative models, statistical inference methods
based on belief propagation were predicted to be optimal in
detecting planted hidden configurations \cite{sbmoptbp,LFASBM,zdeborova2011quiet}. Spectral
methods have the great advantage of being non-parametric, meaning that
they do not require any knowledge of the generative model (if one
exists) and are completely independent of it. Nonetheless, standard
spectral methods
such as the adjacency matrix or the Laplacian succeed down to the
information theoretical limit when the graph is sufficiently dense or
regular \cite{Bickel,vilenchik,coja_partitioning,mcsherry,nadakuditi}
while tend to fail when graphs are sparse due to their sensitivity to
fluctuations in the vertex degree.

These problems have been well studied in the case of graphs
with simple edges between couples of vertices. However, many networks have a
different structure, and the relationships between vertex-variables
are not established in couples but in $k$-uplets, with $k>2$. An
exemple is given, for instance, by the network of scientific
collaborations, of skype conference calls, email exchanges or
recommendation systems where we can associate a user with a specific
content and a rating.  Translating the hypergraph into pairwise
interaction would inevitably lead to a loss of information, and
therefore some effort has been made to generalize spectral methods to
multi-body interactions \cite{dukkipati,zhou,agarwal}.

Here we study an extension of the spectral clustering method 
proposed in \cite{SpRed} based on a generalization of the non-backtracking matrix \cite{hashimoto,SpRed} to the case of {\it hypergraphs} (or factor graphs), that we argue to be effective on sparse networks. 
To test the performance of the spectral method we study it on a
generative stochastic block model of hypergraphs
similar to the ones defined in \cite{abbe13,dukkipati} which is relevant 
for different problems ranging from community detection to planted constraint 
satisfiability. 
We compare the results with statistical inference based on belief propagation, which we also derive, and show the intimate connections 
between the latter and the non-backtracking operator. We also illustrate that on the 
sparse networks we consider, other spectral methods based on standard operators 
fail in the detection where the method we propose succeeds. 

A particularly remarkable point about the method is that it works
without any prior knowledge of the generative model or its parameters, it is
hence a promising tool to learn the probabilistic rules that were
used to create the hypergraph. We illustrate this on the example of
planted constraint satisfaction problems where information about the nature of the
constraints is not assumed but inferred. 

The paper is organized as follows: In Sec. \ref{sec:algo} we give
the form of the generalized non-backtracking matrix and summarize the
algorithm. In Sec. \ref{sec:model} we present the generative
model. In Sec. \ref{sec:spectrum} we then discuss the performance
of the algorithm on hypergraphs generated by the model. In Sec.
\ref{sec:bp} we derive the belief propagation algorithm and the
detectability phase transition by linearization around the uniform
fixed point.  In Sec. \ref{sec:hsbm} and \ref{sec:2in4} we apply
the spectral algorithm to two specific cases of the generative model,
namely an assortative stochastic block model and the planted
$2$-in-$4$-sat, comparing its behavior and phase transitions with the
one of belief-propagation.  Finally in Sec. \ref{sec:conc} we give
our conclusions.

\section{Spectral Detection Algorithm}
\label{sec:algo}
Consider a hypergraph $G(V,E)$, with vertices $V$ with $|V|=N$, and
hyperedges $E$, $|E|=M$. We denote by $\partial \mu$ the set of vertices included
in the hyperedge $\mu$. Each vertex $i$ has an associated variable $a_i$
that can take values in the set ${\cal A}=\lbrace 0, \cdots, q-1
\rbrace$. These variables are hidden from us, but it is assumed that
the hyperedges were chosen in a way that depends on the values of
these variables. The goal is to infer the variables from the structure
of the hypergraph. 

\begin{figure}
\includegraphics[width=.9\columnwidth]{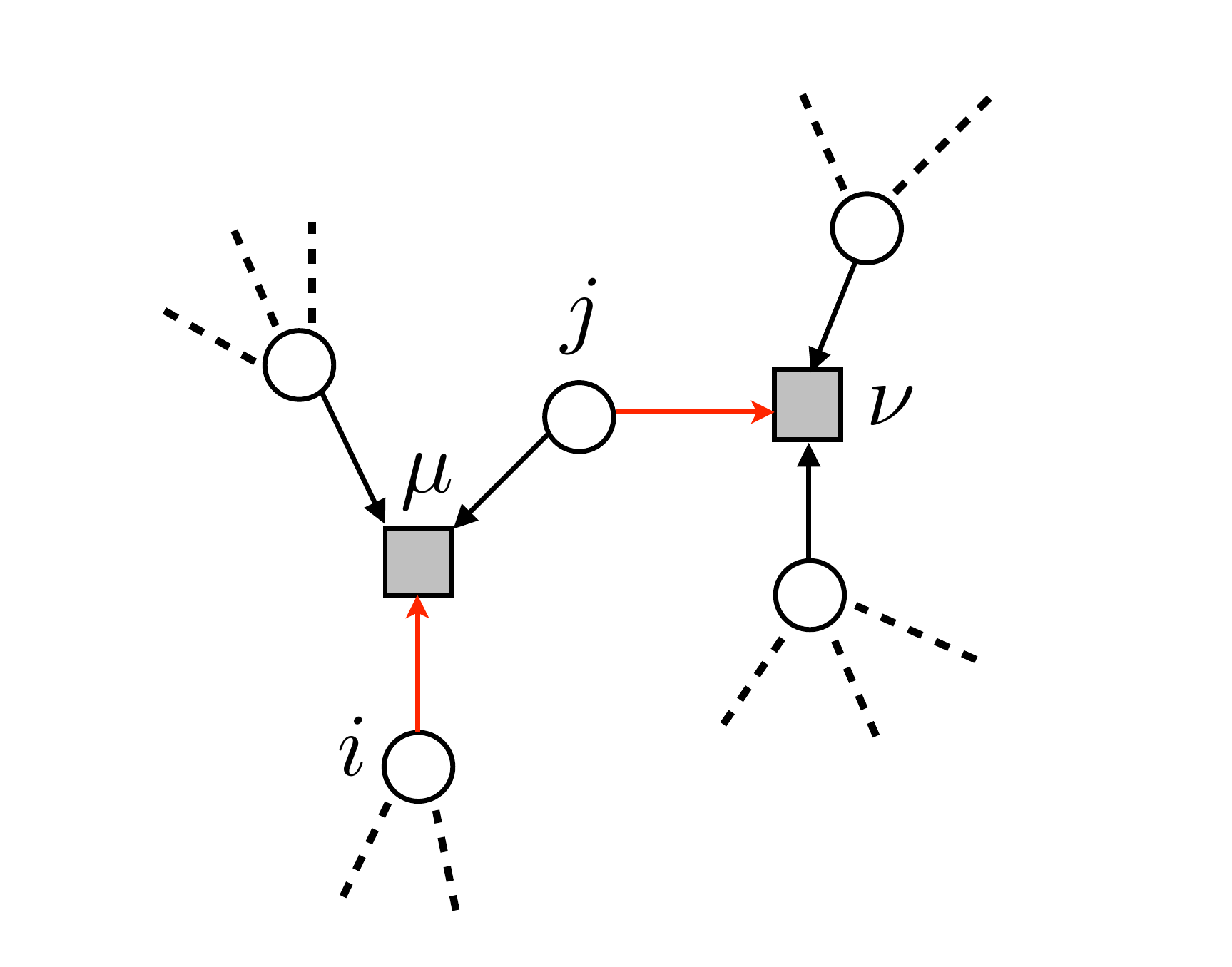} 
\caption{A graphical representation of one non-zero element $B_{(i\to\mu)(j\to\nu)}$ 
of the generalized non-backtracking matrix, where grey squares represent hyper-edges (or factors) and white circles are vertices.}
\label{fig:factor}
\end{figure}

The spectral algorithm we propose to detect hidden labels  
of the vertices in a sparse hypergraph is based on the following hypergraph
non-backtracking operator
\beq
B_{(i\to\mu)(j\to\nu)}=
\begin{cases}
1 & \text{if  } j \in \partial \mu \backslash i \,\, , \,\, \nu\neq
\mu\, , \\
0 & \text{otherwise}\,  ,
\end{cases}
\label{eq:nbo}
\eeq
where $i,j=1,\cdots, N$ are vertex indices and $\mu,\nu=1,\cdots, M$ are hyper-edges (or factors).
In Fig. \ref{fig:factor} we show a graphical representation of one
non-null element of the matrix. Vertices are represented by circles
and hyperedges by squares. Each hyper-edge can be seen as a group of $k_{\mu}$ edges going 
from the participating vertices to the factor $\mu$. 
The matrix is therefore of size $\ol{k} M \times \ol{k} M$, where $M$ is the 
number of factors and 
\beq
\ol{k}=\frac{1}{M}\sum_{\mu} \, k_{\mu} 
\eeq
is the average degree of a factor. 
This matrix has
multiple advantages with respect to the use of adjacency or incidence
matrix, in particular the fact that it is {\it non-backtracking}
inhibits the eigenvalues linked to high degree vertices and the
bulk of the spectrum is confined in a circle in the
complex plane for large random networks. 

The algorithm is the following: 
\begin{itemize}
\item Given the hypergraph, construct the non-backtracking matrix according to (\ref{eq:nbo}).
\item Compute the largest norm eigenvalues (and associated eigenvectors) down to the first that has a non-zero imaginary part. In this way we obtain $q$ eigenvalue-eigenvector pairs. Given that the first eigenvector is associated to 
the degree of the vertices, we retain the subsequent $q-1$ to identify a partition in $q$ groups. 
\item For each of the $q$ eigenvectors ${\bf v}$ construct an $N$ component  vector ${\bf u}$ as 
\beq
u_i=\sum_{\mu\in \partial i} v_{i\to \mu} \, ,
\eeq
where $\partial i$ indicates the set of factors to which vertex $i$ participates. 
\item Run the preferred clustering algorithm, for example soft
  $k$-means, on the components of the
$q-1$ eigenvectors ${\bf u}$ to 
obtain a partition of the vertices in $q$ groups. 
\end{itemize}
The algorithm we propose can be applied to any kind of 
hypergraphs with any distribution of vertex and factor degrees. 

The other spectral method we consider for comparison is based on the adjacency matrix $A$, whose elements $a_{ij}$ are the number of hyper-edges containing both the vertices $i$ and $j$. The adjacency matrix of a hypergraph can be written in terms of the incidence matrix $H$ as $A=H H^{T} - D$, where the incidence matrix is an $N\times M$ matrix such that 
\beq
H_{i \mu}=
\begin{cases}
1 \quad \quad i \in \partial \mu\, , \\
0 \quad \quad {\rm otherwise}.
\end{cases}
\eeq 
In the following we specify a generative model 
on which we will analyze the performance of our non-backtracking algorithm comparing it with 
the ground truth and other algorithms. 

\section{The Model}
\label{sec:model}
Consider a set of vertices $V$ with $|V|=N$, where each vertex variable $a_i$ can 
take values in the set ${\cal A}=\lbrace 0, \cdots, q-1 \rbrace$. Each vertex is independently assigned 
a value $a\in {\cal A}$ with a probability $n_a$ such that $\sum_a n_a=1$.

Consider now a kernel function ${\mathcal P}: {\cal A}^k \to {\rm
  I\!R}_{[0,1]}$ that is symmetric under any permutation of the
arguments and represented by the symmetric $k$-tensor ${\bf p}$
(although generalization to non-symmetric kernels 
should not present any particular difficulty).
Let ${\cal E}^{(k)}$ be the set of all the possible hyper-edges of degree $k$ between the $N$ vertices, the 
kernel tensor gives the probability of existence, independently, for any hyper-edge 
 in ${\cal E}^{(k)}$. To every hyper-edge in ${\cal E}^{(k)}$ is associated an indicator variable $e_{\mu}$ with $\mu = 1, \cdots, \binom{N}{k}$, with probability 
\beq
{\rm Prob}\left(e_{\mu} =1| \, \ul{a}_{\mu}\right)=p_{\ul{a}_{\mu}} \, ,
\eeq
where $\ul{a}_{\mu} \in {\cal A}^k$ is the set of labels planted on the vertices 
participating to the hyper-edge $\mu$. 
The hypergraph is fully specified by 
the vector ${\bf e}$. The expected number of edges in the hypergraph is given by 
\beq
M=\mathbb{E}\left[ \sum_{\mu} e_{\mu} \right]\, .
\eeq
We are interested in the sparse case, i.e. $M=O(N)$, therefore the elements of the tensor must scale as 
\beq
p_{\ul{a}}=\frac{c_{\ul{a}}}{N^{k-1}}\, ,
\eeq
where $c_{\ul{a}}=O(1)$. 
In the following, with a slight abuse of notation, $\ul{a}$ will indicate a variable in 
${\cal A}^k$, ${\cal A}^{k-1}$ or ${\cal A}^{k-2}$, which one of the three will be 
clear in the context.

In the large $N$ limit at leading order the expected degree of a node with 
a label $a$ is 
\beq
c_a=\frac{1}{(k-1)!}\sum_{\ul{b} \in {\cal A}^{k-1}} c_{a,\ul{b}}
\prod_{l=1}^{k-1} n_{b_l} \, .
\eeq
It will also be useful in the following to define the two-vertices average degree, 
namely 
\beq
c_{ab}=\frac{1}{(k-2)!}\sum_{\ul{s} \in {\cal A}^{k-2}} c_{ab,\ul{s}}
\prod_{l=1}^{k-2} n_{s_l}\, . \label{c_ab}
\eeq
Given the planted assignment $\lbrace a_i \rbrace$, the conditional probability of 
generating a certain hypergraph $G$ specified by the set of indicator variables $\lbrace e_{\mu} \rbrace$ is  
\beq
P(G|\lbrace a_i \rbrace_{i=1,\dots,N})= \prod_{\mu \in {\cal E}^{(k)}} \left[
  \left(\frac{c_{\underline{a}_{\mu}}}{N^{k-1}}\right)^{e_{\mu}}
  \left( 1-\frac{c_{\underline{a}_{\mu}}}{N^{k-1}} \right)^{1-e_{\mu}}
\right] \, .
\label{eq:pgraph}
\eeq

As anticipated in the introduction, this generative model covers a wide range of 
problems of which we give some examples below. 
\begin{itemize}
\item {\bf Planted Constraint Satisfaction Problems}. Constraint
  satisfaction problems play a crucial role in theoretical and applied
  computer science as well as in engineering and physics due to their
  very general nature. In a CSP we consider a set of $N$ discrete
  variables, typically Boolean, subject to a set of $M$
  constraints. In many of the usually considered cases like $k$-SAT or
  $k$-XORSAT or $k$-in-$2k$-SAT these constraints are all of the same
  type. In $k$-SAT, for example, the OR between $M$ $k$-uples of
  variables (or their negations) must result to TRUE. In a random CSP
  the constraints are thrown at random between groups of variables,
  giving raise to a random graph. The question is if we can find a
  satisfying assignment. In a planted
  CSP
  \cite{achlioptas2008algorithmic,KrzakalaZdeborova09,krzakala2012reweighted,feldman2014subsampled} we throw at random an assignment of the variables
  and then a series of constraints (factors or hyper-edges) that are
  satisfied by the assignment itself. The question is if and how,
  given the graph, we can recover the planted assignment. Planted CSPs
  are covered by the above generative model.
\item {\bf Hypergraph Stochastic Block Models.} The stochastic block model is 
a popular way of generating graphs with a community structure. An ensemble of $N$ vertices is labeled with values from $1$ to $q$ depending on which 
of the $q$ communities they belong to. Given this assignment, edges between 
couples of nodes are thrown at random with a probability (kernel) that depends 
on the labeling of the two nodes that participate to the edge. Two typical choices 
are the {\it assortative} case where nodes belonging to the same community 
are more likely to be connected or {\it disassortative} if the case is the opposite. 
Again, the objective is, given the graph, recover the underlying community structure. Our model is the natural generalization of the stochastic block model 
to the case of hypergraphs, which is relevant in many applications, from 
recommendation systems to co-authoring networks.
\item {\bf Coding.} If we allow, in the planted CSP scenario, the constraints to be {\it soft}, meaning that 
they can be violated with a certain finite probability, the hypergraph can be seen as a noisy observation of the data, where 
the data consist in the planted assignment. In many cases encoding is performed by summing a random 
set of variables and transmitting the sum through a noisy channel. The choice of the sets results in a 
random hyper-graph with a structure that is determined by the code construction and the transmission gives 
a noisy version of this hypergraph. Our generative model can also be seen as a representation of this kind 
of setting.
\end{itemize}
For the analysis of this paper we will consider the case where the
average degree of the vertices is independent of the labeling, namely
$c_a=c$. This is the statistically hardest case, because simply observing the degree of a node does not give any information about its planted variable. 
In doing so we obtain a random sparse hypergraph with structure but in
which degrees of all the vertices come from Poisson distribution with average $c$.


The performance of a detection algorithm can be evaluated through a measure of the normalized overlap between the planted assignment and the inferred one, namely 
\beq
Q= \max_{\pi}\frac{\frac{1}{N}\sum_i \d_{a_i, \pi(\hat{a}_i)}-\max_a
  n_a}{1-\max_a n_a} \label{overlap}
\eeq
where ${\bf a}$ is the planted assignment, $\hat{{\bf a}}$ is the inferred assignment and $\pi$ is any permutation of the labels.

\section{Properties of $B$ on the planted model}
\label{sec:spectrum}
In this section we present the properties of the non-backtracking
matrix for a hypergraph generated with the above planted model. 

First we remark that the hypergraph can be seen as a bipartite graph
between the variable nodes and the hyperedges. In the case this
bipartite graph is random (non-planted, but fixed degree sequence),
results derived for the spectrum of the non-backtracking
operator of this bipartite graph in \cite{SpRed,Lelarge15} translate directly to the
present case. 

The largest eigenvalue of $B$ is, asymptotically in the large $N$ limit, 
equal to the average branching factor of the locally tree-like
hypergraph. For a $k$-regular hypergraph this reads 
\beq
\mu_1=(k-1)\mathbb{E}_{Q}\left[ d \right]\, , 
\eeq 
where $Q_d=(d+1) p_{d+1}/c$ is the excess degree distribution and $p_d$ is 
the degree distribution. Since we are considering Poissonian degree, 
for the first eigenvalue we find 
\beq
\mu_1= c(k-1)\, .
\eeq
The bulk of the spectrum is confined in the circle of radius 
\beq
\r=\sqrt{c(k-1)}\, .
\label{eq:radius}
\eeq
This can be seen by considering that for any matrix $B$ with 
eigenvalue $\mu$ 
\beq
\sum_{i=1}^{kM} |\mu|^{2r} < \Tr (B^r)(B^r)^{T}\, .
\eeq
Moreover, for any fixed $r$ and in the limit $N\to \infty$, the hypergraph 
is locally tree-like, therefore the diagonal elements $(i\to \mu)(i\to \mu)$ of $(B^r)(B^r)^{T}$ count the exact number of factor nodes at $r$ steps from $\mu$ through paths 
not including $i$, which in expectation is $(k-1)^r c^r$. Therefore for the trace we obtain
\beq
\mathbb{E}\left[\Tr (B^r)(B^r)^{T}\right]=kM(k-1)^r c^r\, ,
\eeq
which gives
\beq
\mathbb{E}\left[|\mu|^{2r} \right] \leq (k-1)^r c^r\, .
\eeq
Since this relation is true for any fixed value of $r$, we conclude that almost all 
the eigenvalues of the non-backtracking matrix lie in the circle of
radius (\ref{eq:radius}). More refined analysis of \cite{Lelarge15}
leads to the result that for a random hypergraph all but one eigenvalue lie
in that circle. 

The planted model can be seen as a perturbed rank-$r$
matrix. If the rescaled eigenvalues of the non-perturbed rank $r$ matrix fall
outside of the circle confining the bulk then they are also visible on
the real axes in the spectrum of the non-backtracking operator. Hence,
in analogy with \cite{SpRed,Lelarge15} the second $q-1$ eigenvalues, when exceeding the bulk, are associated to 
eigenvectors that are correlated with the planted configuration and take values 
\beq
\mu_2= c(k-1) \l\,,
\eeq
where $\l$ is one of the largest $q-1$ eigenvalues of the matrix
\beq
T_{ab}=n_a\left[ \frac{c_{ab}}{c(k-1)} - 1 \right]\, , \label{Tab}
\eeq
which, as we will see, are degenerate under appropriate conditions. 

Inference of the planted assignment can be performed through a standard 
clustering algorithm like $k$-means applied to these eigenvectors when the informative eigenvalues exceed the bulk, namely 
\beq
|\mu_2| > \sqrt{c(k-1)}\, .
\eeq
The vectors that we cluster are constructed from the eigenvectors of $B$ by summing all the outgoing edges for each vertex.

The algorithm hence requires to find the leading eigenvalues of a $\ol{k}M\times \ol{k}M$ matrix, which can be very large if the average degree of the nodes is high. 
Nonetheless, if the hypergraph is $k$-regular as we are considering, we can reduce 
the size of the problem.  
The eigenvalue equation we are solving is the following
\beq
\sum_{\nu \in \partial i \bs \mu }\sum_{k\in \partial \nu \bs i}
v_{k\to \nu}=\mu v_{i\to \mu}\, .
\label{eq_eig}
\eeq
Let us consider the sums of incoming and outgoing messages, respectively
\beq
\begin{split}
v^{\rm in}_{i}&=\sum_{\nu \in \partial i}\sum_{k\in \partial \nu \bs
  i} v_{k\to \nu}\, ,\\
v^{\rm out}_{i}&=\sum_{\nu \in \partial i} v_{i\to \nu}\, ,
\end{split}
\eeq
for which the eigenvalue equation (\ref{eq_eig}) translates into
\beq
\begin{split}
(d_i-1) v^{\rm in}_{i}&=\l v^{\rm out}_{i}\, ,\\
\sum_{\mu \in \partial i}\sum_{j\in \partial \mu \bs i} v^{\rm in}_{j}
- (k-2) v^{\rm in}_{i} - (k-1) v^{\rm out}_{i} &= \l v^{\rm in}_{i}\, .
\end{split}
\eeq
The preceding equation can be written in a compact form as
\beq
B'{\bf v}=\l {\bf v}\, ,
\eeq
with
\beq
{\bf v}=\left(\begin{matrix} {\bf v}^{\rm out} \\  {\bf v}^{\rm
      in}\end{matrix}\right)\, ,
\eeq
and
\beq
B'=\left(\begin{matrix} 0 & D-\mathbb{1}\\ -(k-1)\mathbb{1} &
    A-(k-2)\mathbb{1}\end{matrix}\right)\, ,
\eeq
where $B'$ is a $2N\times 2N$ matrix, $D$ is the diagonal degree matrix $A$ is the symmetric adjacency matrix.
Given this, we see that all the eigenvalues of the complete non-backtracking operator $B$ are also eigenvalues of the reduced one $B'$ except for those $k M - 2N$ associated to the subspace defined by 
\beq
\begin{split}
\sum_{\nu \in \partial i}\sum_{k\in \partial \nu \bs i} v_{k\to \nu}=0
\quad \forall i\, ,\\
\sum_{\nu \in \partial i} v_{i\to \nu}=0 \quad \quad \forall i\, ,
\end{split}
\eeq 
for which we cannot assure the correspondence.
Moreover, by a further transformation of the eigenvalue eq. (\ref{eq_eig}), we can obtain the 
following non-linear equation describing the eigenvalues of the reduced operator
\beq
v_i= \frac{1}{1-\frac{\l(p-2+\l)}{p-1}} \left[ d_i v_i - \frac{\l}{p-1} \sum_{\mu \in \partial i} \sum_{k \in \partial \mu \bs i} v_k \right]
\eeq
telling us that the differences between the $B$ and $B'$ spectra will be located in $\l=1$ and $\l=-(p-1)$ due 
to the singularities that appear in the formula.

\begin{figure*}[ht]
(a)
  \includegraphics[width=.62\columnwidth]{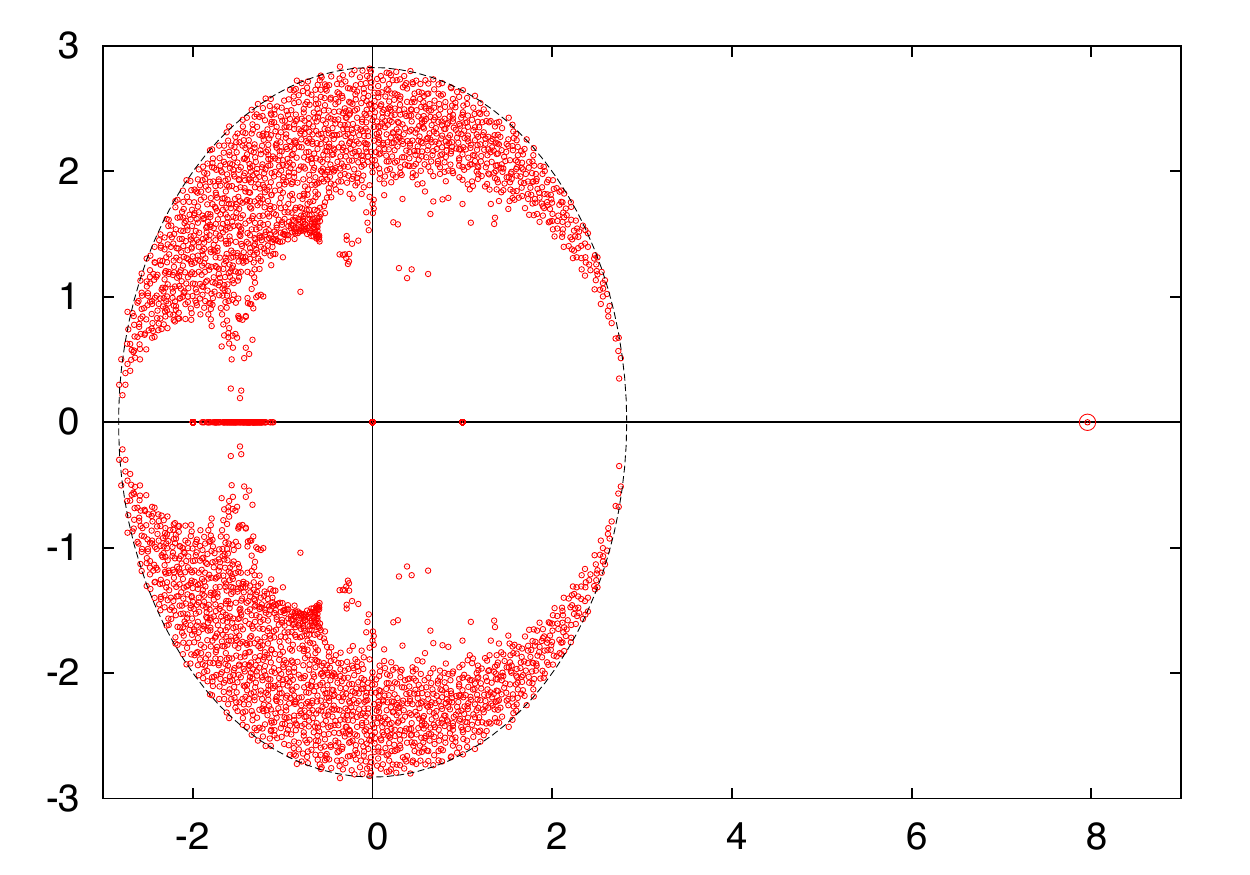}
(b) 
 \includegraphics[width=.62\columnwidth]{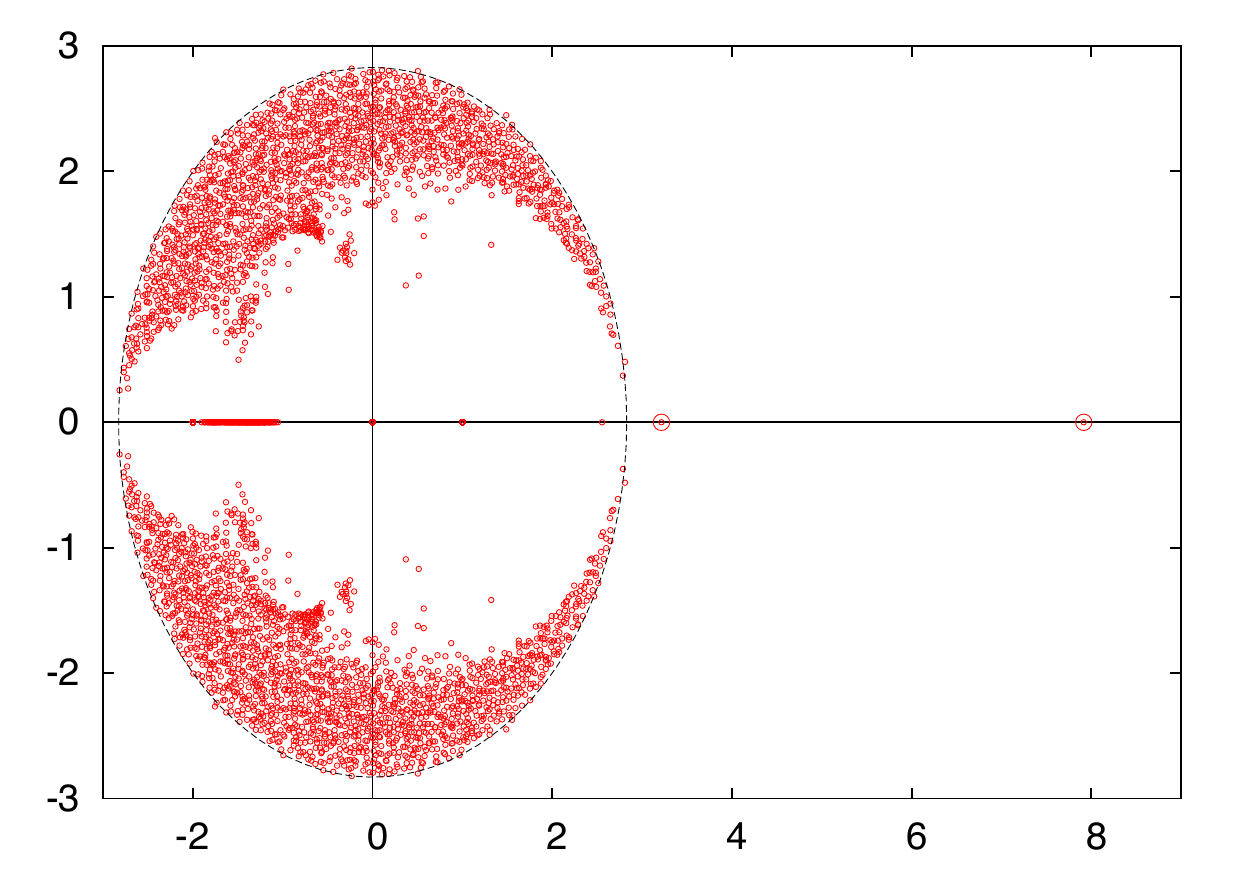} 
(c) 
\includegraphics[width=.6\columnwidth]{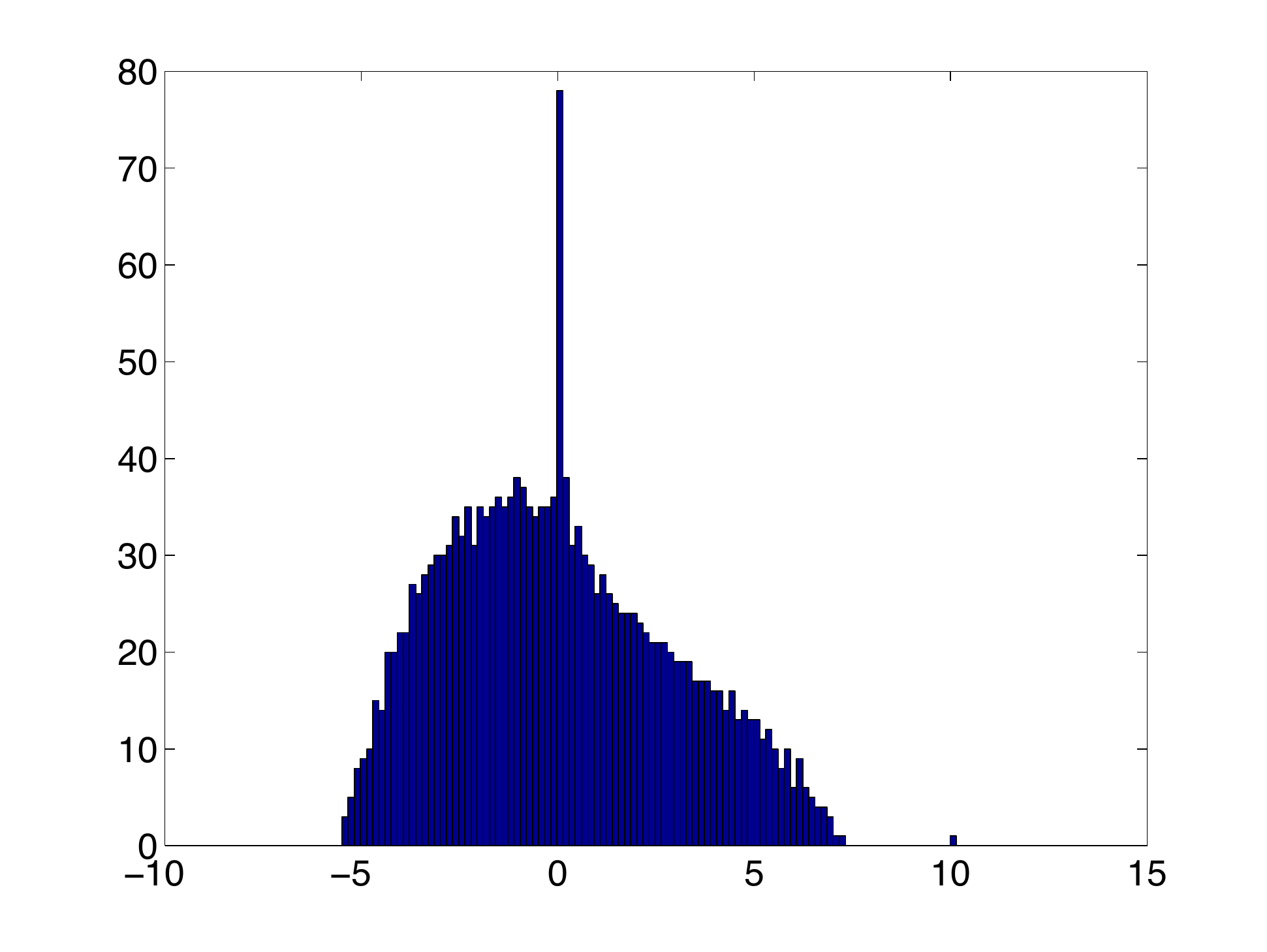} 
\caption{The spectrum of the hypergraph non-backtracking matrix
  (\ref{eq:nbo}) for an hypergraph of size $N=1800$ generated through
  the assortative HSBM with $k=3$, $q=3$ and $c=4$ for (a)
  $\tilde{\e}=0.22$ and (b) $\tilde{\e}=0.14$. In (b) the informative
  eigenvalues lie outside the bulk and are degenerate. (c) The
  spectrum of the adjacency matrix for $\tilde{\e}=0.14$. Despite being in the detectable phase the informative eigenvalue of the
  adjacency matrix is hidden in the bulk.}
\label{fig:specassort}
\end{figure*}

\section{Belief Propagation and Phase Transitions}
\label{sec:bp}
In this section we derive the asymptotic properties of the generative model through belief propagation and, as a byproduct, 
we highlight the connection between belief propagation and the generalized non-backtracking operator.

Given the conditional probability of a graph (\ref{eq:pgraph}),
according to the Bayes rule the posterior probability 
of the planted assignment ${a_i}$ given the hypergraph $G$ reads
\beq
\begin{split}
P(\lbrace a_i \rbrace_{i=1,\dots,N} | G)&=\frac{1}{Z(G)} \prod_{i=1}^N n_{a_i}  \times \\
&\times \prod_{\mu \in {\cal E}^{(k)}} \left[
  c_{\underline{a}_{\mu}}^{e_{\mu}} \left(
    1-\frac{c_{\underline{a}_{\mu}}}{N^{k-1}} \right)^{1-e_{\mu}}
\right]\, ,
\end{split}
\label{eq:pplant}
\eeq
where $Z(G)$ is the proper normalization constant. This posterior probability is associated 
to a graphical model for which we can write $2k|{\cal E}^{(k)}|$ coupled message-passing equations \cite{MezardMontanari09}
\beq
\begin{split}
\psi_{a}^{\mu \to i}& =\frac{1}{Z^{\mu \to i}} \sum_{\ul{b}} \left(
  \prod_{j \in \partial \mu \backslash i} \chi^{j\to \mu}_{b_j}\right)
\times\, , \\
&\times \left[ c_{a,\underline{b}}^{e_{\mu}} \left(
    1-\frac{c_{a,\underline{b}}}{N^{k-1}} \right)^{1-e_{\mu}}
\right]\, ,\\
\chi^{i\to \mu}_{a}&=\frac{1}{\Theta^{i \to \mu}} \, n_a \, \prod_{\nu
  \in {\tilde \partial} i \backslash \mu}  \psi_{a}^{\nu \to i}\, ,
\end{split}
\label{eq:mess}
\eeq
where ${\tilde \partial}i$ indicates all the hyperedges in ${\cal E}^{(k)}$
that contain vertex $i$, $Z^{\mu \to i}$ and $\Theta^{i \to \mu}$ are
normalization constants. The 
estimated marginal probability that a vertex was planted in the group $a$ is given by the associated belief, namely 
\beq
\chi^{i}_{a}=\frac{1}{\Theta^{i}} \, n_a \, \prod_{\nu \in {\tilde \partial}
  i}  \psi_{a}^{\nu \to i}\, .
\eeq
By plugging the first equation of (\ref{eq:mess}) into the second, we obtain a closed set of $k|{\cal E}^{(k)}|$ equations for the $\chi$ messages:
\beq
\begin{split}
\chi^{i\to \mu}_{a}&=\frac{1}{K^{i \to \mu}}\, n_a \, \prod_{\nu \in {\tilde \partial} i \backslash \mu} \sum_{\ul{b}} \left( \prod_{j \in \partial \nu \backslash i} \chi^{j\to \nu}_{b_j}\right) \times \\
&\times \left[ c_{a,\underline{b}}^{e_{\nu}} \left(
    1-\frac{c_{a,\underline{b}}}{N^{k-1}} \right)^{1-e_{\nu}}
\right]\, .
\end{split}
\eeq
In order to reduce the number of equations we note that there are two different types of messages $i\to \mu$, namely the ones living on the $kM$ edges that actually exist in $G$, 
for which $i\in \partial \mu$ and $e_{\mu}=1$, and the ``ghost'' edges for which $i\notin \partial \mu$ and $e_{\mu}=0$.

It can be shown that, up to corrections that vanish when $N\to \infty$, for 
the ``ghost'' hyper-edges we have $\chi^{i\to \mu}_a \simeq \chi^i_a$, while for the messages living on the real edges 
\beq
\chi^{i\to \mu}_a=\frac{e^{-h_a}}{K^{i \to \mu}} \, n_a \prod_{\nu
  \in \partial i \backslash \mu} \sum_{\ul{b}} \left( \prod_{j
    \in \partial \nu \backslash i} \chi^{j\to \nu}_{b_j}\right) \,
c_{a,\ul{b}}\, ,
\label{eq:messpass}
\eeq
with 
\beq
h_a=\frac{1}{N^{k-1}} \sum_{\g \in {\cal E}^{(k-1)}} \sum_{\ul{b}}\,
c_{a,\ul{b}} \, \left( \prod_{j \in \partial \g} \,\chi^{j}_{b_j}
\right)\, ,
\eeq
which reduces the number of messages to $kM$. Still, the computation of the effective field $h_a$ requires 
the summation of order $N^{k-1}$ terms which for large hypergraphs becomes problematic as soon as $k>2$. We will 
show in Sec. \ref{sec:appli} that the computation of the field can be largely simplified when a specific choice of the 
connectivity tensor ${\bf c}$ is made. 

It can be easily verified that, if the condition $c_a=c \,\, \forall a$ holds, then the so-called factorized 
solution $\chi^{i\to \mu}_a=n_a$ is a fixed point of eq. (\ref{eq:messpass}). 
In order for the inference of the planted assignment to be easy, the above mentioned factorized fixed point must be unstable, 
guaranteeing that belief propagation does not remain trapped in it.

To analyze the linear stability of the factorized fixed point under random 
perturbation of the messages. Let us consider messages of the form
\beq
\chi^{i\to \mu}_a= n_a + \epsilon^{i\to \mu}_a\,.
\label{eq:pert}
\eeq
Plugging (\ref{eq:pert}) into (\ref{eq:messpass}) and developing to first order we obtain
\beq
\epsilon^{i\to \mu}_a = \sum_{b} \, T_{ab} \, \sum_{j\in \partial \nu
  \setminus i} \, B^T_{(i\to \mu)(j\to\nu)} \epsilon_b^{j\to\nu}\, ,
\eeq
where $T_{ab}$ is the matrix from (\ref{Tab}) and $B$ is the
hypergraph non-backtracking matrix.

The hypergraphs generated by the model in the regime we are considering are sparse and locally tree-like, meaning that on average loops start to be observed at a distance $O(\log N)$. Let us then consider a tree of depth $d$ and observe 
how a perturbation on a leaf propagates through the (unique) path connecting it to the root 
\beq
\d \chi_{a_0}^0=\sum_{a_1,\cdots, a_d} \, \left[\prod_{i=0}^{d-1} \,
  T_{a_i a_{i+1}}\right]\d\chi_{a_d}^d=\sum_{a_d}(T^d)_{a_0
  a_{d}}\d\chi_{a_d}^d\, .
\eeq
Now, taking independent random perturbations, summing up the contributions of all the leaves and considering that 
in the limit $d\to \infty$ the matrix $T^d$ is dominated by its largest eigenvalue $\l$, we have 
\beq
\d {\bf \chi}^0 \simeq \sum_{k \in {\rm leaves}} \l^d \d {\bf
  \chi}_{||}^k\, ,
\eeq
where $\d {\bf \chi}_{||}^k$ is the perturbation along the direction of the dominating eigenvector.
Therefore, in terms of expectation we obtain
\beq
\mathbb{E}\left[\d\chi^0\right]=0\, ,
\eeq
and
\beq
\mathbb{E}\left[\left(\d\chi^0\right)^2\right]\simeq c^d (k-1)^d
\l^{2d} \mathbb{E}\left[\left(\d\chi_{||}\right)^2\right]\, .
\eeq
The instability threshold is finally given by 
\beq
c (k-1) \l^{2}=1\, .
\eeq
When $c (k-1) \l^{2}<1$ the factorized fixed point is stable and belief
propagation will hence not be able to infer the planted assignment,
and for $c (k-1) \l^{2}>1$ belief propagation succeeds.

Moreover, if we restrict ourselves 
to the case $n_a=1/q$ with $c_{ab}$ from (\ref{c_ab}) being
\beq
c_{ab}=
\begin{cases}
c_{\rm in} & \text{if   }  a=b\, , \\
c_{\rm out} &\text{if   } a\neq b\, .
\end{cases} \label{cinout}
\eeq
and consequently
\beq
T_{ab}=
\begin{cases}
T_{\rm in} & \text{if   }  a=b \, ,\\
T_{\rm out} &\text{if   } a\neq b\, ,
\end{cases}
\eeq
then we find the two following eigenvalues
\begin{eqnarray}
&\l_1 &= T_{\rm in}+ (q-1) T_{\rm out}=0\, ,\\
&\l_2 &= T_{\rm in}-  T_{\rm out}=\frac{c_{\rm in}-  c_{\rm
    out}}{q(k-1) c}\, ,
\end{eqnarray}
with multiplicities of $1$ and $q-1$ respectively. Therefore the instability has a closed expression, namely
\beq
\frac{|c_{\rm in}-  c_{\rm out}|}{q}=\sqrt{c(k-1)}\, ,
\label{eq:instab}
\eeq
with the expected degree given by
\beq
c=\frac{c_{\rm in}+(q-1) c_{\rm out}}{q(k-1)}\, .
\eeq

\section{Examples}
\label{sec:appli}

\subsection{Hypergraph Stochastic Block Model}
\label{sec:hsbm}
Restricting ourselves to the case (\ref{cinout}) and by analogy with the $k=2$ case \cite{LFASBM,SpRed}, we 
define the hypergraph stochastic block model (HSBM) as assortative if $c_{\rm in}>c_{\rm out}$ and disassortative otherwise. 
A sensible choice for the parametrization of the problem is by $\e=c_{\rm out}/c_{\rm in}$ and the expected degree $c$. 
With this notation the transition is located at
\beq
\e_c=\frac{\sqrt{c(k-1)}-1}{\sqrt{c(k-1)}+(q-1)} \, .
\eeq
In this section we consider the specific case 
\beq
c_{a_1 \cdots a_k}=
\begin{cases}
\tilde{c}_{\rm in} \quad {\rm if} \quad a_1=a_2=\cdots = a_k\, ,\\
\tilde{c}_{\rm out} \quad {\rm otherwise}\, ,
\end{cases}
\eeq
for which we can define the specific parameter 
\beq
\tilde{\e}=\tilde{c}_{\rm out}/\tilde{c}_{\rm
  in}=\frac{\e}{q^{k-2}+(q^{k-2}-1)\e}\, .
\eeq
As anticipated, with this setting the computation of the effective field in the belief propagation iteration takes linear time and simplifies 
in the following way up to $O(1/N)$ corrections
\beq
h_a= \tilde{c}_{\rm out} + \frac{(\tilde{c}_{\rm in} -
  \tilde{c}_{\rm out})}{(k-1)! \, N^{k-1}} \left(\sum_{i=1}^N \chi^i_a
\right)^{k-1} + O\left(\frac{1}{N}\right)\, .
\eeq
We take as an example a HSBM with $k=3$, $q=3$ and $c=4$ which gives
the detectability transition located at $\tilde{\e}_c= 0.1688$. In
Fig. \ref{fig:specassort} we show the spectrum of the non-backtracking
operator (a) in the undetectable phase (where the factorized belief propagation fixed point is
stable) and (b) in the detectable phase (where belief propagation gives an informative fixed
point). While in the undetectable phase we find only the leading eigenvalue associated to the 
average excess degree, in the detectable phase two more eigenvalues stick out of the bulk and the correspondent eigenvectors are correlated with the community structure. 

\begin{figure}
\includegraphics[width=.9\columnwidth]{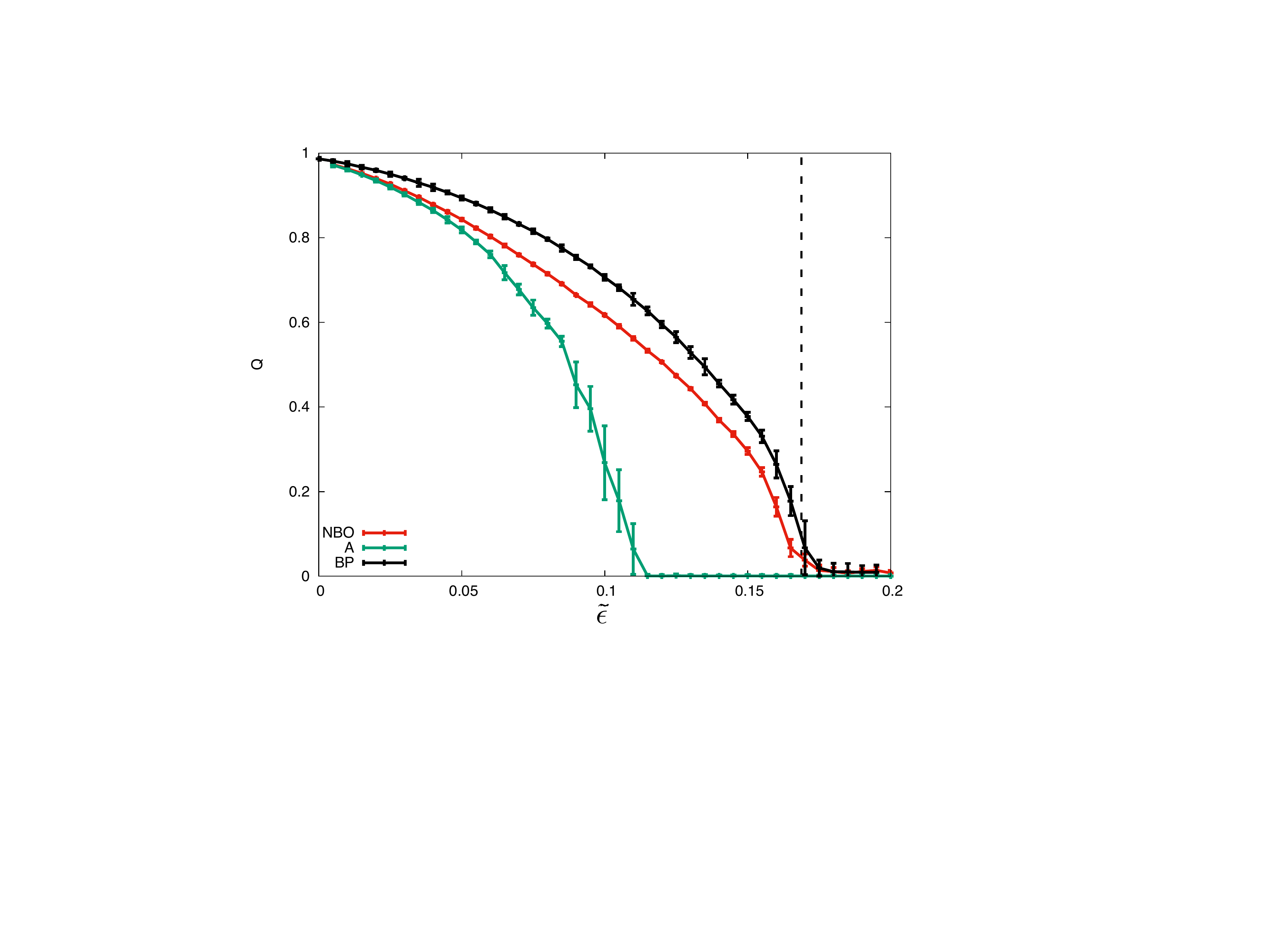} 
\caption{Performance, in terms of the overlap $Q$ (\ref{overlap}), of
  the spectral detection on the assortative HSBM through the
  non-backtracking matrix (NBO) compared to spectral detection with
  the adjacency matrix (A) and Bayesian belief propagation (BP). The size of the graph is $N=300000$ for the three curves and averages are taken over $5$ samples. The vertical dashed line marks the detectability transition.}
\label{fig:assortperf}
\end{figure}

In Fig. \ref{fig:assortperf} we show the performance of the spectral clustering through the non-backtracking operator 
combined with a standard $k$-means. Despite a slightly worse
performance, it displays the same phase transition as
belief-propagation, which we conjectured to be an optimal algorithm \cite{LFASBM}. By contrast, spectral clustering through the adjacency matrix 
has a comparable performance deep in the detectable phase but it breaks down well before the phase transition due to the sparsity of the hypergraph.

\begin{figure*}[ht]
(a)
  \includegraphics[width=.62\columnwidth]{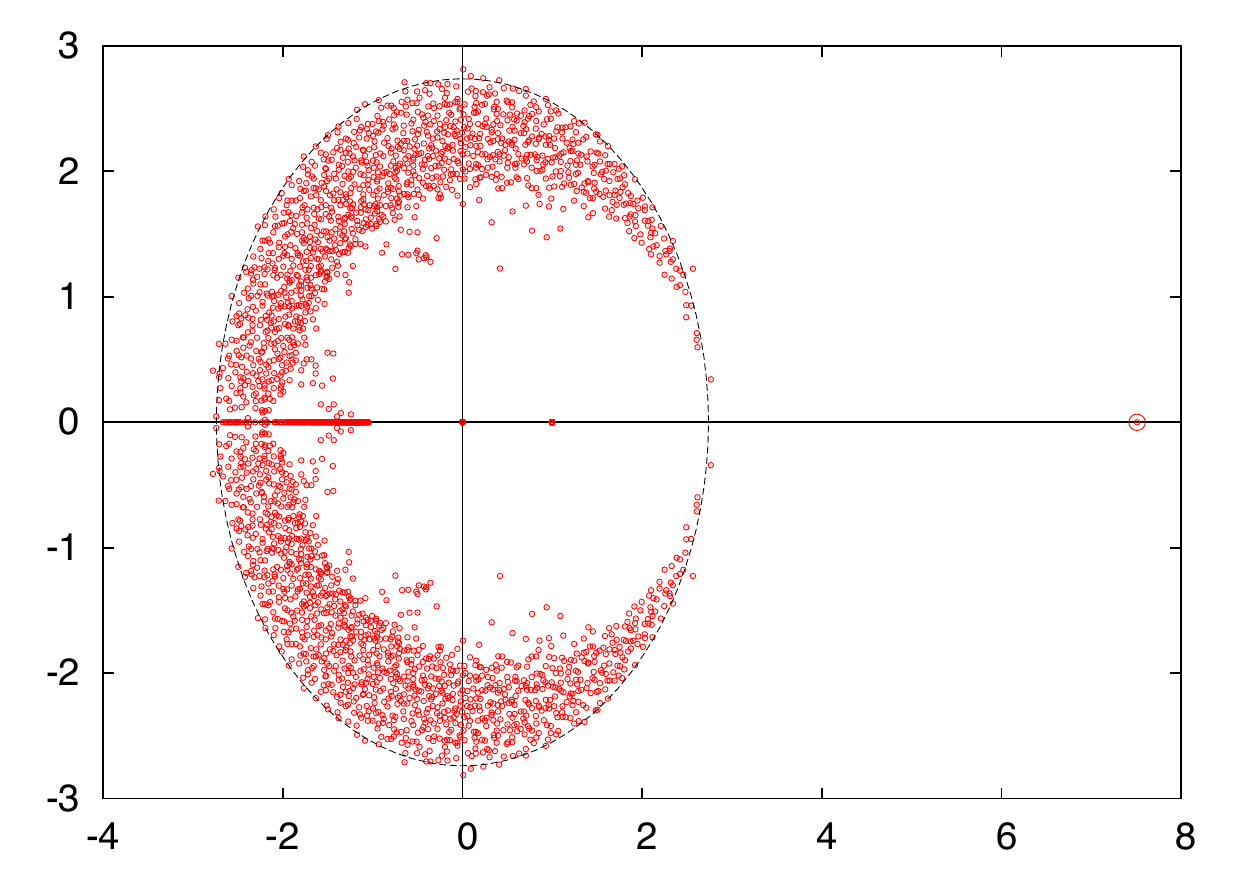}
(b)
 \includegraphics[width=.62\columnwidth]{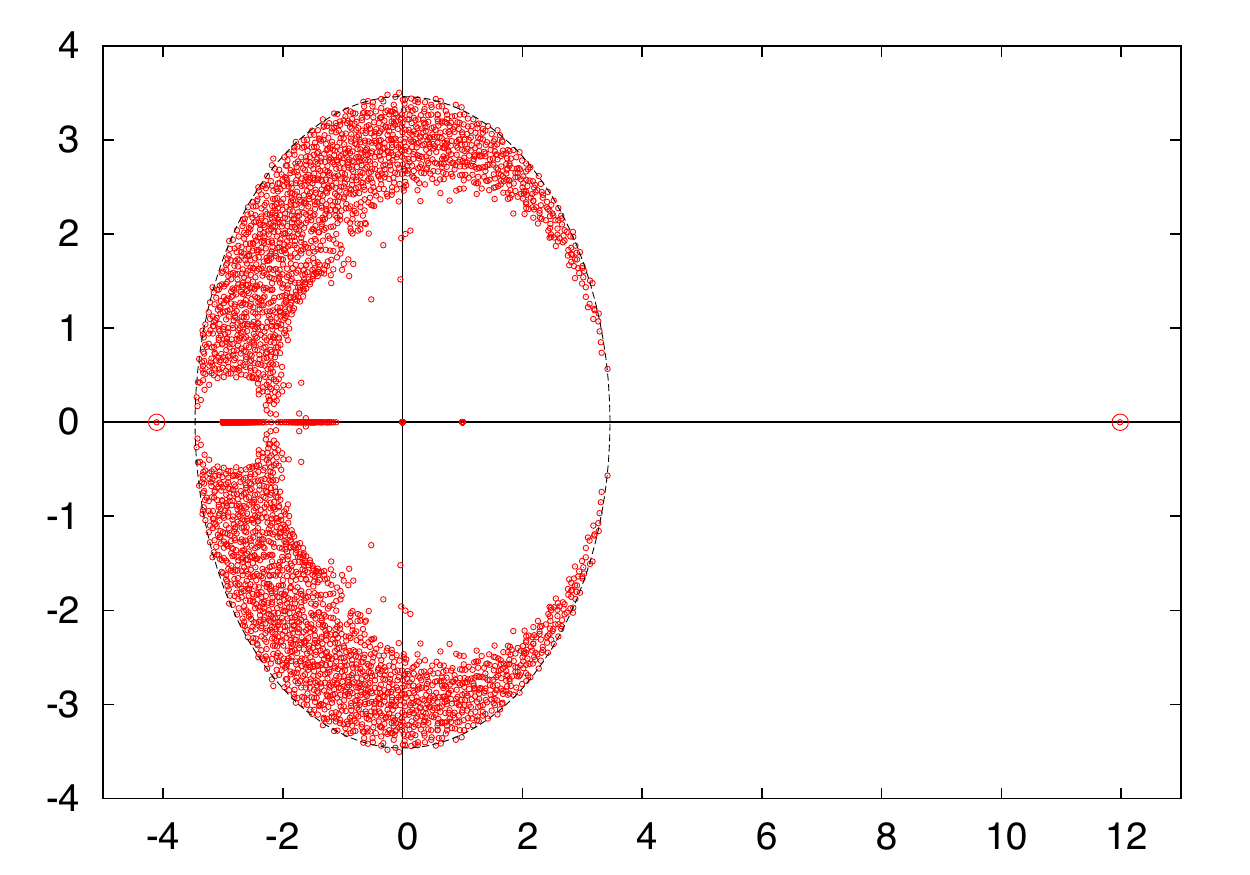}
(c)
\includegraphics[width=.6\columnwidth]{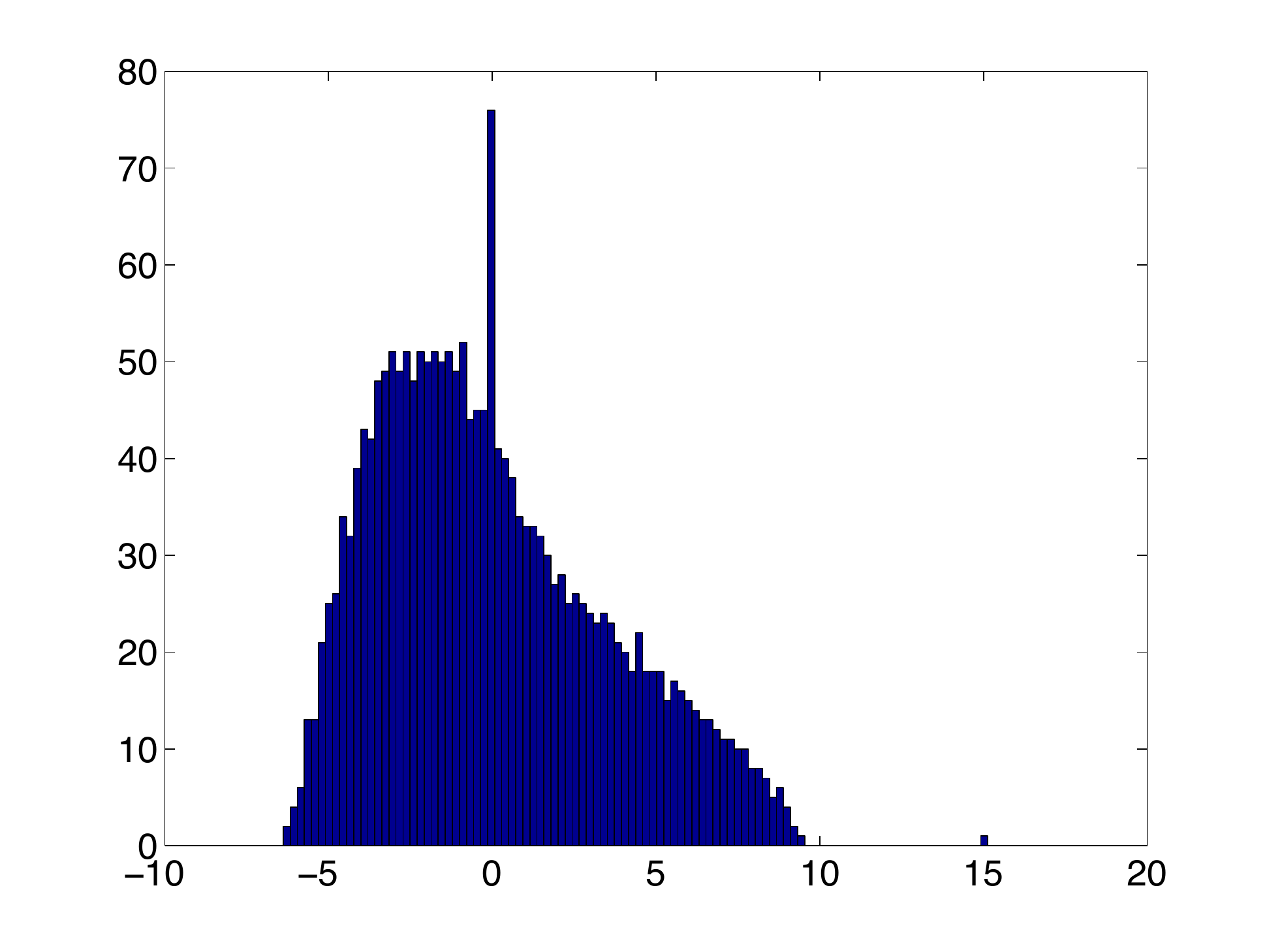} 
\caption{(a,b) The spectrum of the hypergraph non-backtracking matrix (\ref{eq:nbo}) for a hypergraph 
of size $N=2000$ generated through the planted $2$-in-$4$-sat ($k=4$, $q=2$)  for (a) $c=2.5$ and (b) $c=4$. In (b) the informative 
eigenvalue lie outside the bulk on the left because the model is disassortative. In (c) the spectrum of 
the adjacency matrix for $c=4$. The informative eigenvalue is hidden in the bulk.}
\label{fig:spec2in4}
\end{figure*}

\subsection{Planted $2$-in-$4$-sat}
\label{sec:2in4}
In the planted $2$-in-$4$-sat problem, after giving a random binary assignment to the 
vertices of the network, hyperedges are thrown at random between groups of $k=4$ vertices with a 
certain non-zero probability only if in the $k$-uplet there is an equal number of zeros and ones. 
At fixed expected degree $c$, this generative process fits into our general formulation with 
\beq
\begin{split}
c_{0011}&=16c\, ,\\
c_{\rm in}&=2c\, ,\\
c_{\rm out}&=4c\, .
\end{split}
\label{eq:2in4inout}
\eeq This problem has been studied through the cavity method in the so
called ``locked'' case in \cite{zdeb_locked}. Here we are interested
in the problem of finding a configuration correlated to the planted
one.  The scenario is quite different with respect to the preceding
case. In fact here we find two phase transitions in the average
degree, namely one from an impossible to hard detection
\cite{zdeborova2011quiet,zdeb_locked}, and a second one from hard to
easy detection that is our main focus. In the hard phase we conjecture that no known
algorithm is able to retrieve the planted assignment due to the
presence of the stable factorized fixed point, despite the fact that
the global fixed point would be the one at high overlap.  In this case
the hard/easy phase transition in belief propagation is discontinuous
and the transition is located at \beq c_c=3\, , \eeq which is obtained
by plugging (\ref{eq:2in4inout}) into (\ref{eq:instab}).  In this case
the computation of the effective field in the belief propagation
equations is reduced to linear time with \beq h_a\simeq \frac{8c}{N^3}
\, \left(\sum_{i=1}^N \chi_a^i\right)\left(\sum_{i=1}^N
  \chi_{1-a}^i\right)^2 + O\left(\frac{1}{N}\right)\, .  \eeq Since
this planted problem is essentially disassortative, as shown in
Fig. \ref{fig:spec2in4} the informative eigenvalue sticks out at the
left of the spectrum.

\begin{figure}
\includegraphics[width=.9\columnwidth]{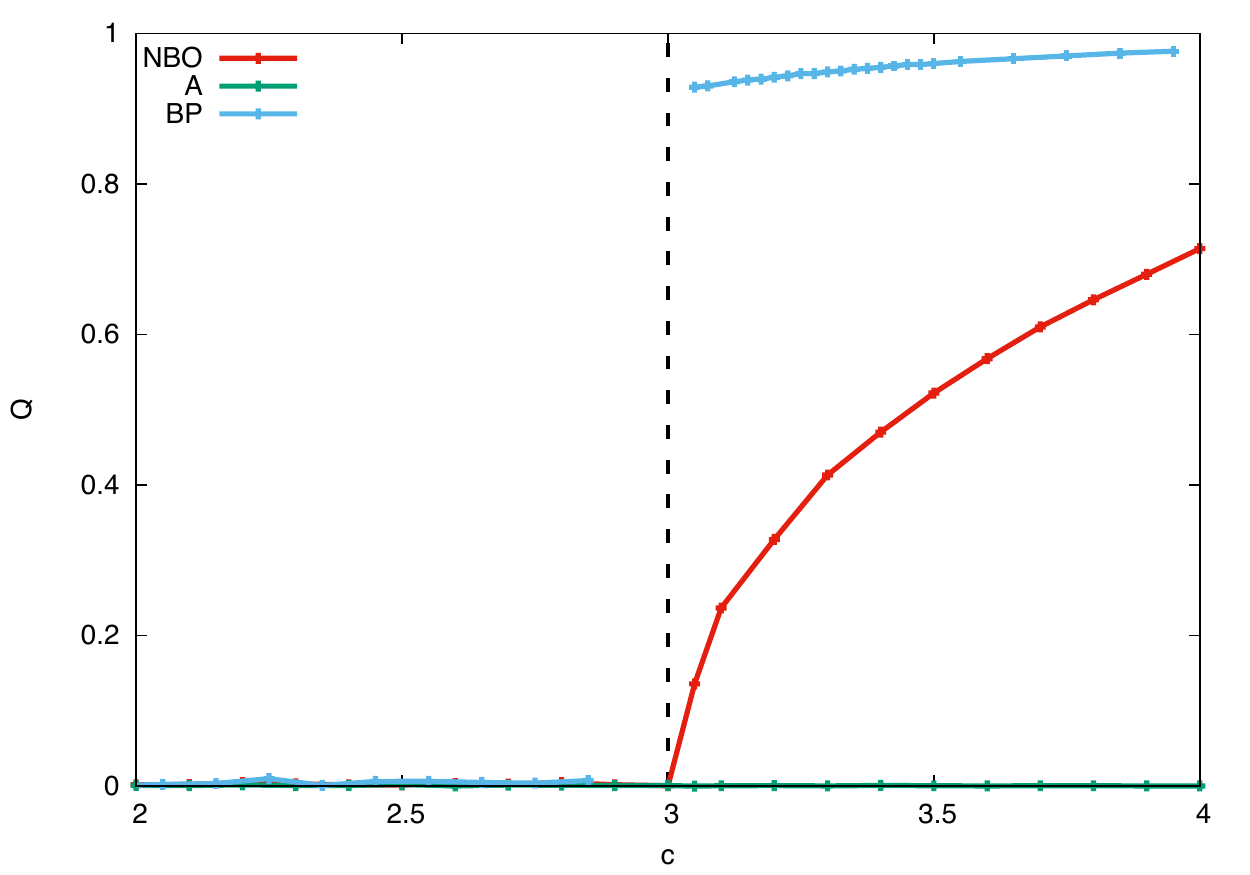} 
\caption{Performance, in terms of the overlap $Q$ (\ref{overlap}), of
  the spectral detection on the planted $2$-in-$4$-sat through the
  non-backtracking matrix (NBO) compared to spectral detection with
  the adjacency matrix (A) and Bayesian belief propagation (BP). The
  size of the graph is $N=200000$ for  belief propagation and
  $N=400000$ for the spectral algorithms. In both cases a single
  sample was taken. The vertical dashed line marks the hard/easy
  transition. The adjacency matrix starts giving a non-zero overlap
  only at very high average degree, namely 
around $c\approx 9$ which is out of the plot range.}
\label{fig:2in4perf}
\end{figure}

In Fig. \ref{fig:2in4perf} we show the performance of the spectral algorithm based on the generalized non-backtracking matrix compared with spectral clustering on the adjacency matrix and with belief propagation. The adjacency matrix displays a terrible behavior up to 
very large average degree. The non-backtracking matrix instead undergoes
a phase transition, located at the same value of the one we encounter
in belief-propagation but the transition is continuous. Despite the fact that the non-backtracking operator performs 
considerably worse than belief propagation, the important point to underline is that the spectral method is completely non-parametric and we 
do not even need to know what the kernel is. In fact, we can think of the spectral method also as a way to {\it learn} the kernel, in order to feed it as a starting condition for the hyper-parameters in belief propagation with parameter learning. 
Let us take as an example a graph generated with $c=3.4$ and $N=40000$, after running the spectral detection we have 
an inferred labeling of each vertex. Then let us take the list of $M=34000$ factors and look at their composition (according to the inferred labels). The result is shown in Table \ref{tab}, telling us that we are likely observing a planted $2$-in-$4$-sat model or something very close to it. 

\begin{table}[h]
\centering
\caption{2-in-4 kernel estimation}
\label{tab}
\begin{tabular}{|l|l|}
\hline
0000 &  1/34000\\ \hline
0001 &  5764/34000\\ \hline
0011 &  21758/34000\\ \hline
0111 &  6477/34000\\ \hline
1111 &  0/34000 \\ \hline
\end{tabular}
\end{table}

\section{Conclusion}
\label{sec:conc}
In this paper we have proposed a spectral algorithm to detect hidden
planted configurations in very sparse hypergraphs, based on a
generalization of the non-backtracking Hashimoto matrix.  To test the
performance of the algorithm we have focused on a generative
probabilistic model for hypergraphs in which the hyperedges depend on
the incident variables via a fixed probability kernel.  Given the
generative model, we have also derived an asymptotically (conjectured)
optimal belief propagation algorithm and presented a derivation of the
non-backtracking matrix as a linearization of belief propagation equations around the
factorized fixed point. In addition we have used belief propagation to
compute the location of the detectability phase-transition.

The generative model that we consider includes a broad class of
problems. Among them we have studied the assortative stochastic block
model and the planted $2$-in-$4$-sat. In the first case we obtained
that the spectral non-backtracking clustering has a performance that
is very close to the optimal belief propagation and displays a
detectability phase transition at the same point. It also performed
much better than the spectral clustering based on the adjacency matrix
that breaks down way before the phase transition due to the sparsity
of the graph.  In the second case of the planted $2$-in-$4$-sat we
observed a different phenomenon, reminiscent of a first order
phase transition. While for belief propagation the phase transition is discontinuous
from a hard inference phase to an easy inference phase with the
overlap that jumps from zero to a value close to one, in the spectral
detection algorithm we observe a continuous transition at the very
same point. Spectral detection with the adjacency matrix, again,
performs badly up to even higher degree. In both cases, the gain
provided by the non-backtracking approach was clear.

We also showed that despite the fact that the accuracy of the spectral
method is significantly worse w.r.t. belief propagation (although they
start to detect assignments at the same values of the parameters), the
spectral approach has many important advantages: not only it is entirely
non-parametric, but it is also a powerful instrument to learn the
parameters, such as the kernel, when they are unknown.



\bibliographystyle{unsrt}
\bibliography{refs}

\end{document}